\documentclass[a4paper,12pt]{article}
\usepackage{a4,epsfig}

\newcommand{\sgn}{\mathrm{sgn}}
\newcommand{\Cos}{\mathbf{C}}
\newcommand{\Sin}{\mathbf{S}}
\newcommand{\tr}[1]{\mathrm{tr}\left\{ #1 \right\}}
\newcommand{\ket}[1]{\left|{#1}\right\rangle}
\newcommand{\bra}[1]{\left\langle{#1}\right|}
\newcommand{\adj}{^{\dagger}}
\newcommand{\rhoss}{\varrho^{\mathrm{(ss)}}}
\newcommand{\rhoA}{\varrho^{\mathrm{(1st)}}}
\newcommand{\rhoB}{\varrho^{\mathrm{(2nd)}}}
\newcommand{\rhoAB}{\varrho^{\mathrm{(both)}}}
\newcommand{\sep}{\varrho^{\mathrm{(both)}}_{\mathrm{sep}}}
\newcommand{\pure}{\varrho^{\mathrm{(both)}}_{\mathrm{pure}}}
\newcommand{\DoS}{\mathcal{S}}
\newcommand{\Drho}{\Delta\varrho}
\newcommand{\Nex}{N_{\mathrm{ex}}}
\newcommand{\half}{\frac{1}{2}}
\newcommand{\fourth}{\frac{1}{4}}
\newcommand{\Vec}[1]{\vec{\,#1}}
\newcommand{\Dyad}[1]{\raisebox{\depth}{%
${{\scriptscriptstyle\longleftrightarrow} \atop {\displaystyle #1}}$}}

\begin{document}
\thispagestyle{empty}

\begin{center}
\Large
\textbf{One-atom maser: Non-separable atom pairs}

\bigskip\bigskip
\large
Berthold-Georg Englert, Pavel Lougovski, Enrique Solano,\\ 
and Herbert Walther

\medskip

\normalsize
Max-Planck-Institut f\"ur Quantenoptik\\ 
Hans-Kopfermann-Str.~1, 85748~Garching, Germany

\bigskip\bigskip

\centerline{(13 September 2002)}

\bigskip\bigskip

\parbox{0.9\textwidth}{%
\centerline{\textbf{Abstract}}
When a one-atom maser is operated in the standard way 
--- excited, resonant two-level atoms 
traverse the resonator at random times --- 
the emerging atoms are entangled with the cavity field.
As a consequence, the results of measurements on different atoms tend to be
correlated.
We show that truly non-classical correlations can be found between two
successive atoms by studying the properties of the reduced state 
of such an atom pair.
In particular, we calculate its degree of separability and find parameter
ranges in which it is markedly less than unity.
}

\vfill

Dedicated to the memory of Professor Aleksandr Mikhailovich \textsc{Prokhorov}.

\end{center}

\newpage

\section{Introduction}
When a two-level atom and a photon mode undergo a Jaynes--Cummings
interaction \cite{JC}, they become entangled as a rule. 
Various aspects of atom-field entanglement of this kind have been studied in
past years, whereby the emphasis is usually --- but not always \cite{GHZ+EPR}
--- on the consequent atom-atom entanglement that results from the interaction
of several atoms with the same photon mode.

Clearly, if two atoms interact with the photons at the same time, rather
strong atom-atom entanglement is expected, and this is indeed the case (see,
e.g., the recent paper by Masiak \cite{pairs}).
For similar reasons one also gets strong entanglement if particular
initial internal states of the atoms are suitably combined with well chosen
interaction times (see, e.g., \cite{GHZ+EPR,imperfectBell}).

Much less obvious are the properties of the atom-atom entanglement that is
generated in standard one-atom maser (OAM) experiments \cite{Fortschritte}.
In general terms it is known that there are correlations among emerging atoms,
that they manifest themselves in non-Poissonian counting statistics 
\cite{Rempe90}, for instance, and that measurements on consecutive atoms 
can be inconsistent with a Bell-type inequality~\cite{pseudoBell}.

But a systematic study of the properties of the joint state of two atoms that
have emerged from the resonator of a OAM experiment has not been performed
beyond the extraction of some correlation data of experimental 
relevance~\cite{Fortschritte}.
It is the objective of the present contribution, which is a progress report in
character, to contribute to this systematic study.

We shall here focus on what is perhaps the simplest situation in this context,
namely two atoms where the second traverses the resonator very shortly after
the first leaves. 
One anticipates correctly that such atom pairs exhibit particularly strong
entanglement because the dissipative degradation of the photon field 
can be disregarded for the brief period between the two atoms.

\section{Successive atoms in a one-atom maser} 
To set the stage, we first recall some basics of OAM theory 
(see, e.g., \cite{Fortschritte}).
The resonant Jaynes--Cummings dynamics evolves the joint state 
of a two-level atom 
(excited state: $\ket{\mathsf{e}}$, ground state: $\ket{\mathsf{g}}$)
and the photon field 
(state of definite photon number: $\ket{n}$, with $n=0,1,2,\ldots$)
in accordance with
\begin{eqnarray}
  \label{eq:A1}
\ket{\mathsf{e},n,\ldots}&\to& \Cos\ket{\mathsf{e},n,\ldots}
                             -i\Sin\ket{\mathsf{g},n,\ldots}
\nonumber\\   
  &=&\ket{\mathsf{e},n,\ldots}\cos\left(\varphi\sqrt{n+1}\,\right)
-i\ket{\mathsf{g},n+1,\ldots}\sin\left(\varphi\sqrt{n+1}\,\right)\,, 
\end{eqnarray}
where the ellipsis indicates quantum numbers that are not affected (such as
those referring to other atoms) and
\begin{equation}
  \label{eq:A2}
  \Cos=\cos\left(\varphi\sqrt{aa\adj}\,\right)\,,\qquad
  \Sin=a\adj\frac{\sin\left(\varphi\sqrt{aa\adj}\,\right)}{\sqrt{aa\adj}}
\end{equation}
are essentially trigonometric functions of the photon ladder operators $a$ and
$a\adj$.
The accumulated Rabi angle $\varphi$ is the net measure of the interaction 
strength and duration.

Prior to the arrival of the first atom, 
the photon field is in the steady state,
specified by the statistical operator $\rhoss$ that is given by \cite{FJM}
\begin{equation}
  \label{eq:A3}
  \rhoss(a\adj a)=\rhoss(0)\prod_{n=1}^{a\adj a}
  \left[\frac{\nu}{\nu+1}+\frac{\Nex}{\nu+1}
              \frac{\sin^2\bigl(\varphi\sqrt{n}\,\bigr)}{n}\right]\,.
\end{equation}
Here, the normalized pump rate $\Nex$ is the mean number of atoms traversing
the resonator in one photon lifetime, $\nu$ is the number of thermal
photons in the cavity field, and the value of $\rhoss(0)$ is determined by the
normalization of $\rhoss$ to unit trace.

The final state of the first atom is thus given by
\begin{equation}
  \label{eq:A4}
  \rhoA=\ket{\mathsf{e}}\tr{\Cos\rhoss\Cos}\bra{\mathsf{e}}
        +\ket{\mathsf{g}}\tr{\Sin\rhoss\Sin\adj}\bra{\mathsf{g}}
       =\half(1+\Vec{s}\cdot\Vec{\sigma})\,,
\end{equation}
where
\begin{equation}
  \label{eq:A5}
  \Vec{s}=(0,0,s)\quad\mbox{with}\quad s=\tr{(\Cos^2-\Sin\adj\Sin)\rhoss}
\end{equation}
is the Bloch vector for the first atom and
\begin{equation}
  \label{eq:A6}
  \Vec{\sigma}=(\sigma_x,\sigma_y,\sigma_z)
    =\Bigl(\ket{\mathsf{g}}\bra{\mathsf{e}}+\ket{\mathsf{e}}\bra{\mathsf{g}},
      i\ket{\mathsf{g}}\bra{\mathsf{e}}-i\ket{\mathsf{e}}\bra{\mathsf{g}},
      \ket{\mathsf{e}}\bra{\mathsf{e}}-\ket{\mathsf{g}}\bra{\mathsf{g}}\Bigr)
\end{equation}
is the Pauli vector operator of the familiar pseudo-spin formalism for
two-level atoms.

For the second atom, we denote the Pauli vector operator
by $\Vec{\tau}$ and the numerical Bloch vector by $\Vec{t}$.
Since the second atom is assumed to arrive immediately after the first has
left --- this is the ``perhaps simplest situation'' alluded to in the
Introduction --- its statistical operator is given by
\begin{equation}
  \label{eq:A7}
  \rhoB=\half(1+\Vec{t}\cdot\Vec{\tau})
\end{equation}
with
\begin{equation}
  \label{eq:A8}
\Vec{t}=(0,0,t)\,,\quad
  t=\tr{(\Cos^2-\Sin\adj\Sin)(\Cos\rhoss\Cos+\Sin\rhoss\Sin\adj)}\,.  
\end{equation}
Rare exceptions aside, the values of $s$ and $t$ will be different, which is,
of course, a sign that the cavity field ``remembers'' that the first atom has
just passed through the resonator.
Photon dissipation makes this memory fade in the course of time and,
therefore, it is quite obvious that the correlations between two successive
atoms are strongest if the second comes immediately after the first.
In the present paper, we are dealing with this particularly favorable 
situation.
Results about more general cases will be reported elsewhere. 

Now turning to the correlated statistics of two successive atoms, 
we note that,
in addition to the Bloch vectors $\Vec{s}$ and $\Vec{t}$, 
the statistical operator $\rhoAB$ of the joint two-atom state,
\begin{equation}
  \label{eq:A9}
  \rhoAB=\fourth\left(1+\Vec{s}\cdot\Vec{\sigma}+\Vec{t}\cdot\Vec{\tau}
                       +\Vec{\sigma}\cdot\Dyad{C}\cdot\Vec{\tau}\right)\,,
\end{equation}
involves the cross dyadic $\Dyad{C}$, which happens to be diagonal,
\begin{equation}
  \label{eq:A10}
\Dyad{C}=\left(\begin{array}{ccc}u&0&0 \\ 0&u&0 \\ 0&0&v \end{array}\right)
\end{equation}
with
\begin{eqnarray}
  \label{eq:A11}
  u&=&\tr{(\Sin\adj\Cos\Sin\Cos+\Cos\Sin\adj\Cos\Sin)\rhoss}
\nonumber\\\mbox{and}\quad 
  v&=&\tr{(\Cos^2-\Sin\adj\Sin)(\Cos\rhoss\Cos-\Sin\rhoss\Sin\adj)}\,.
\end{eqnarray}
By construction, all eigenvalues of $\rhoAB$ are non-negative, which is to say
that the four parameters $s$, $t$, $u$, and $v$ obey the inequalities
\begin{equation}
  \label{eq:A11a}
  1-v\geq\sqrt{4u^2+(s-t)^2}\,,\qquad 1+v\geq|s+t|\,.
\end{equation}

The difference between $\rhoAB$ and the two-atom product state $\rhoA\rhoB$,
\begin{equation}
  \label{eq:A12}
  \Drho=\rhoAB-\rhoA\rhoB
       =\fourth\Vec{\sigma}\cdot\Dyad{E}\cdot\Vec{\tau}\,,
\end{equation}
is measured in terms of the entanglement dyadic 
$\Dyad{E}=\Dyad{C}-\Vec{s}\Vec{t}$. 
Here its only non-zero elements are the values $u,u,v-st$ on the diagonal.
Their absolute values are the characteristic values of $\Dyad{E}$, and the
largest of them is the maximal ``degree of correlation'' \cite{Fortschritte},
which is one measure of the strength of the entanglement between the two
atoms.

Another and perhaps more telling measure is the absolute size of $\Drho$,
its trace norm, $\tr{\left|\Drho\right|}$, the sum of the moduli of the
eigenvalues of $\Drho$. 
These four eigenvalues are $\fourth(2u-v-st),\fourth(-2u-v-st)$, and twice
$\fourth(v-st)$, so that we get
\begin{equation}
  \label{eq:A13}
  \tr{\left|\Drho\right|}=\left\{
    \begin{array}{ll}
|v-st| &\mbox{if $2|u|\leq|v-st|$,}\\[1ex]
\half|v-st|+|u| &\mbox{if $2|u|\geq|v-st|$,}
    \end{array}\right.
\end{equation}
for this measure of entanglement.

\section{Degree of separability}
Although an entangled two-atom state $\rhoAB$ is, by definition, not equal to
the product $\rhoA\rhoB$ of the two single-atom states contained in it, it may
very well be a convex sum of such products.
Then one speaks of a \emph{separable} state,
\begin{equation}
  \label{eq:B1}
 \mbox{$\rhoAB$ is separable if}\qquad \rhoAB=\sum_{k}w_k\rhoA_k\rhoB_k\qquad
\mbox{with $w_k>0$}.
\end{equation}
Entangled, but separable states exhibit correlations that can be modeled
classically: It is as if we had one of the products $\rhoA_k\rhoB_k$ without
knowing which one, the $k$-th occurring with probability $w_k$.

The Peres--Horodeccy criterion \cite{Peres,Hor} makes it easy to check whether
the $\rhoAB$ of (\ref{eq:A9})--(\ref{eq:A13}) is separable for the given 
set of parameter values.
One finds that
\begin{equation}
  \label{eq:B2}
\mbox{$\rhoAB$ is separable if}\qquad 
1+v\geq\sqrt{4u^2+(s+t)^2}
\qquad\mbox{and only then}\,.
\end{equation}
In particular, we have a separable $\rhoAB$ whenever $u=0$.

According to what is said in the context of (\ref{eq:B1}), 
truly non-classical correlations require a non-separable state.
Now, there are non-separable states that differ very little 
from separable ones and others that differ by much. 
A quantitative measure for a non-separable state's proximity to separable ones
is the \emph{degree of separability} $\DoS$, which is defined by 
the Lewenstein--Sanpera decomposition of $\rhoAB$
\cite{LewSan}, 
\begin{equation}
  \label{eq:B3}
  \rhoAB=\DoS\sep+(1-\DoS)\pure\,,
\end{equation}
where $\sep$ is separable, $\pure$ is a pure state, and $\DoS$ is maximal.
The smaller the value of $\DoS$, the stronger are the non-classical
correlations in $\rhoAB$.

For states of the particularly simple 4-parametric structure
(\ref{eq:A9})--(\ref{eq:A13}), one can use known methods \cite{EngMet}
to find the decomposition (\ref{eq:B3}), with this outcome: 
The pure part is of the form
\begin{equation}
  \label{eq:B4}
  \pure=\fourth\left[1+p\sigma_z-p\tau_z
               +q\left(\sigma_x\tau_x+\sigma_y\tau_y\right)
                -\sigma_z\tau_z\right]
\end{equation}
with $q=\sgn(u)\sqrt{1-p^2}$ and the value of $p$ is determined as follows.
We have
\begin{equation}
  \label{eq:B5}
p=0\quad  \mbox{if}\quad \Lambda^2-(1+v)\Lambda+st\geq0
\end{equation}
with
\begin{equation}
  \label{eq:B6}
  \Lambda=1-|u|+\half\sqrt{(1+v)^2-(s+t)^2}
\end{equation}
and get $p$ as the solution of 
\begin{equation}
  \label{eq:B7}
  \frac{1-\Lambda}{\sqrt{1-p^2}}\left[(1-v)-(s-t)p\right]=(1-\Lambda)^2-v+st
\end{equation}
when (\ref{eq:B5}) does not apply.
The degree of separability is then given by
\begin{equation}
  \label{eq:B8}
  \DoS=1-\frac{1-\Lambda}{\sqrt{1-p^2}}\,,
\end{equation}
so that $\DoS=\Lambda$ when $p=0$, and $\DoS<\Lambda$ when $p\neq0$.

\section{Examples and summary}
\begin{figure}[h]
\begin{center}
\epsfig{file=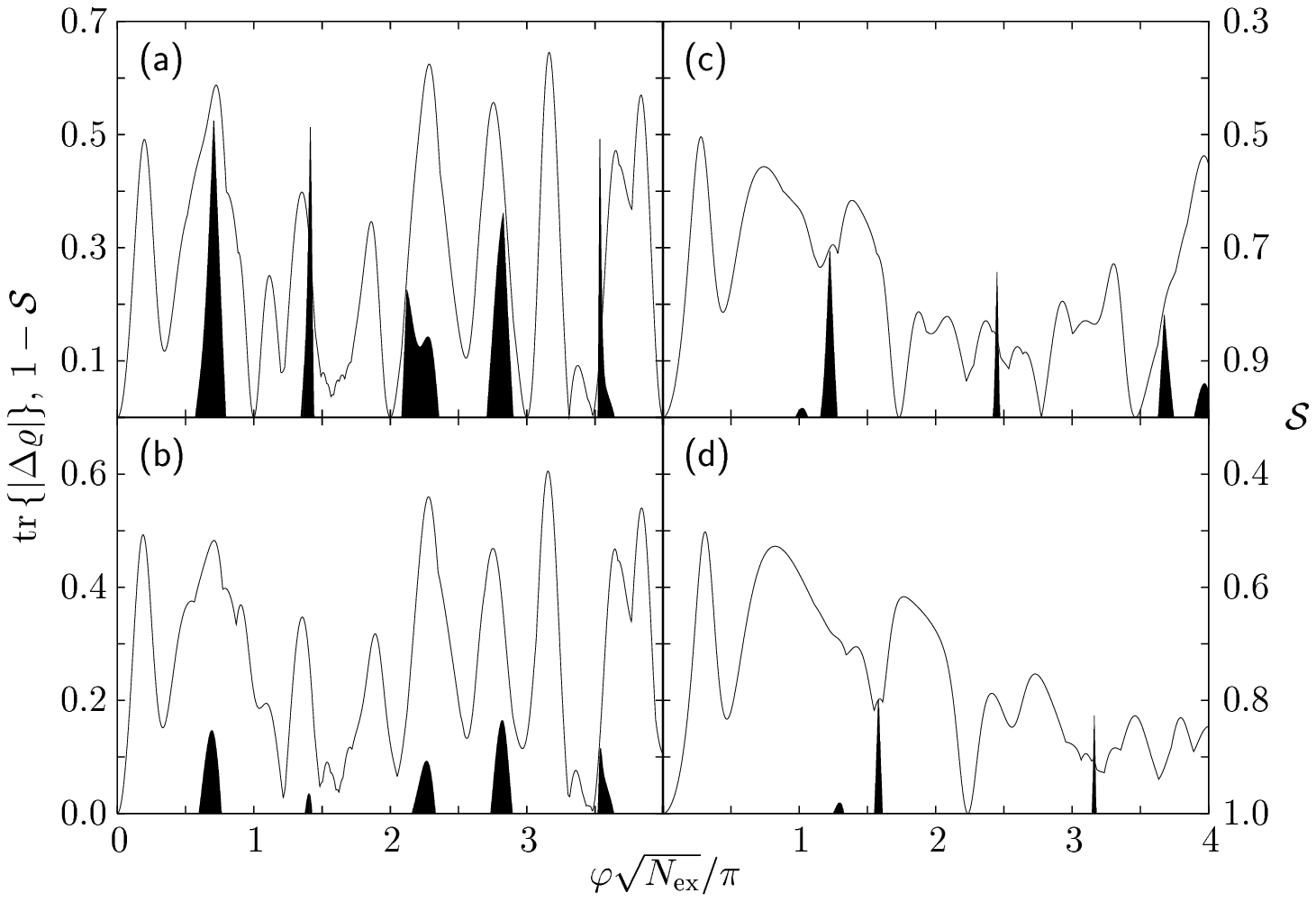,width=0.95\textwidth}
\end{center}    
\caption[Aa]{\label{fig:examps}%
Measures for entanglement and separability of two atoms that traverse the
resonator in immediate succession.
As a function of the ``pump parameter'' $\varphi\sqrt{\Nex}$, the solid line
shows the size $\tr{|\Drho|}$ of $\Drho$ 
[cf.\ (\ref{eq:A12}) and (\ref{eq:A13})], 
and the contour of the dark area shows the degree of separability $\DoS$
[cf.\  (\ref{eq:B3}) and (\ref{eq:B8})]. 
Plot (a) is for $\Nex=1$, $\nu=0$; plot (b) for $\Nex=1$, $\nu=0.2$;
plot (c) for $\Nex=3$, $\nu=0$; and plot (d) for $\Nex=5$, $\nu=0$.
In plot (a), the peak values of $1-\DoS$ at $\varphi=0.708\pi$, $1.414\pi$,
and $3.536\pi$ are $0.5245$, $0.5130$, and $0.4920$, respectively.
} 
\end{figure}

In Fig.~\ref{fig:examps} we show some numerical results obtained for the OAM
parameters $(\Nex,\nu)=(1,0)$, $(1,0.2)$, $(3,0)$, and $(5,0)$.
All are for rather small fluxes, namely $\Nex=1$ in plots (a) and (b),
$\Nex=3$ in plot (c), and $\Nex=5$ in plot (d).
In the present context, small fluxes have the advantage that the steady state
$\rhoss$ does not have a large average photon number, so that the additional
photon that can be deposited by the first atom may lead to a substantial 
change in the cavity field.
The field then encountered by the second atom is markedly different
from $\rhoss$ and, therefore, there is a good chance to generate a strong
entanglement between the atoms.

Put differently, with a large flux, $\Nex\gg1$, one will have a steady state
with a photon number distribution spread over a wide range.
The statistical average over the various values of $n$ in (\ref{eq:A1}) will
then tend to degrade the atom-atom correlations.
Indeed, the comparison of plots (a), (c), and (d) ---
where $\Nex=1$, $3$, and $5$, respectively --- demonstrates that both
$\tr{|\Drho|}$ and $1-\DoS$ have somewhat smaller values for larger $\Nex$.
The effect on $\DoS$ is particularly pronounced as the total dark area is
much smaller in plot (c) than in plot (a), and even smaller in plot (d).

Similarly, one expects that the noise associated with thermal photons 
has a negative influence on the atom-atom correlations.
This is confirmed by the comparison between plots (a) and (b), where $\nu=0$
and $\nu=0.2$, respectively.
Again we notice some reduction in $\tr{|\Drho|}$ and a marked loss of dark
area. 

In summary, then, we have demonstrated that a one-atom maser in standard
operation can give rise to rather strong entanglement between successive
atoms, provided the parameters are chosen judiciously (low flux, low
temperature).
For some parameter ranges, one obtains non-separable two-atom states with
truly non-classical correlations.
 
We have confined the discussion to the simple situation of atoms that traverse
the resonator in immediate succession.
Clearly, one must allow for a more reasonable time gap when aiming at a
comparison with real-life experimental data. 
Work in this direction is in progress and results will be reported in due
course.
Further, the entanglement properties of chains of more than two atoms
are not understood to a satisfactory extent.
So far there have only been studies of bulk quantities, 
such as the mean number of successive atoms in the same 
final state \cite{Fortschritte,same1,same2},
and surely there is a lot of finer detail that deserves more attention.

\section*{Acknowledgments}
BGE's work is supported in part by a TITF initiative of Texas A\&M University.

\newpage

\section*{Figures}

\vfill

\vfill

\end{document}